%% file: main.tex
\documentclass[conference]{IEEEtran}
\IEEEoverridecommandlockouts
\usepackage[style=ieee,backend=biber]{biblatex}
\usepackage[backend=biber]{biblatex}
\usepackage{amsmath,amssymb,amsfonts}
\usepackage{algorithmic}
\usepackage{graphicx}
\usepackage{textcomp}
\usepackage{xcolor}
\usepackage{booktabs}
\usepackage{graphicx}
\usepackage{adjustbox}
\usepackage{multicol}
\usepackage[linesnumbered,algoruled,boxed,lined]{algorithm2e}
\usepackage{fancyhdr}

\fancypagestyle{firstpage}
{
    \fancyhead[C]{\smaller©2021 IEEE. Personal use of this material is permitted. Permission from IEEE must be obtained for all other uses, in any current or future media, including reprinting/republishing this material for advertising or promotional purposes, creating new collective works, for resale or redistribution to servers or lists, or reuse of any copyrighted component of this work in other works.}
    \setlength{\headheight}{27.06pt}
    \fancyfoot{}
}

\SetAlFnt{\small}

\begin{document}

\title{Accelerating JPEG Decompression on GPUs\\
}

\author{\IEEEauthorblockN{André Weißenberger}
\IEEEauthorblockA{\textit{Institute of Computer Science} \\
\textit{Johannes Gutenberg University}\\
Mainz, Germany \\
aweissen@uni-mainz.de}
\and
\IEEEauthorblockN{Bertil Schmidt}
\IEEEauthorblockA{\textit{Institute of Computer Science} \\
\textit{Johannes Gutenberg University}\\
Mainz, Germany \\
bertil.schmidt@uni-mainz.de}
}

\maketitle

\thispagestyle{firstpage}

\begin{abstract}
The JPEG compression format has been the standard for lossy image compression for over multiple decades, offering high compression rates at minor perceptual loss in image quality. For GPU-accelerated computer vision and deep learning tasks, such as the training of image classification models, efficient JPEG decoding is essential due to limitations in memory bandwidth. As many decoder implementations are CPU-based, decoded image data has to be transferred to accelerators like GPUs via interconnects such as PCI-E, implying decreased throughput rates. JPEG decoding therefore represents a considerable bottleneck in these pipelines. In contrast, efficiency could be vastly increased by utilizing a GPU-accelerated decoder. In this case, only compressed data needs to be transferred, as decoding will be handled by the accelerators. In order to design such a GPU-based decoder, the respective algorithms must be parallelized on a fine-grained level.
However, parallel decoding of individual JPEG files represents a complex task. In this paper, we present an efficient method for JPEG image decompression on GPUs, which implements an important subset of the JPEG standard. The proposed algorithm evaluates codeword locations at arbitrary positions in the bitstream, thereby enabling parallel decompression of independent chunks. Our performance evaluation shows that on an A100 (V100) GPU our implementation can outperform the state-of-the-art implementations libjpeg-turbo (CPU) and nvJPEG (GPU) by a factor of up to 51 (34) and 8.0 (5.7). Furthermore, it achieves a speedup of up to 3.4 over nvJPEG accelerated with the dedicated hardware JPEG decoder on an A100.
\end{abstract}


\section{Introduction}
The JPEG standard, as released in 1992 \cite{wallace1992jpeg} has become the most widely used method for lossy image compression. It was adopted in digital photography, and initially enabled the use of high quality photos on the world wide web, most notably in social media. 

In recent years, graphics processing units (GPUs) have emerged as a powerful tool in deep learning applications. For example, specialized hardware like NVIDIA's DGX Deep Learning Systems enable training of image classification models with very high efficiency. These systems consist of host components (multicore CPUs and main memory) and a number of GPU accelerators. Accelerators can communicate with each other via a high-bandwidth interconnect (NVLink) but are connected to the host system via the much slower PCIe bus. 

As various methods of classification are often applied to photographic images, the data is stored mostly in JPEG format. Unfortunately, JPEG bitstreams cannot easily be decoded utilizing fine-grained parallelism, due to data dependencies. Encoded data consists of short bit sequences of variable length. Therefore, the locations of those sequences need to be known prior to decoding when the input is subdivided into blocks for parallelization. For this reason, the process of decoding a batch of images is usually parallelized at a coarse-graind level, e.g., by processing separate files with different CPU threads. This implies a bottleneck for many systems like the DGX: before any image data can be processed by the GPUs, it needs to be decoded on the host side. Uncompressed data, which can be multiple times as large as the compressed data, needs to be transferred to the compute devices via the interconnect afterwards which causes bottlenecks.

For this reason, it would be desirable to perform the necessary decompression on the respective GPU immediately, leading to increased efficiency. nvJPEG is a GPU-accelerated library for JPEG decoding released by NVIDIA, and is part of the Data Loading Library (DALI) for accelerating decoding and augmentation of data for deep learning applications. Ampere-based GPUs (like the A100) even include a 5-core hardware JPEG decode engine to speedup nvJPEG \cite{nvjpeghardware}. For other GPUs, however, nvJPEG employs a hybrid (CPU/GPU) approach and does not entirely run on the GPU.

In this paper, we present a novel design for a fully GPU-based JPEG decoder. By making use of the \textit{self-synchronizing property} of Huffman codes, fine-grained parallelism is enabled, resulting in a more scalable algorithm. We will mainly focus on the parallelization of the bitstream decoding stage of JPEG using CUDA, as the remaining steps necessary to reconstruct the final image have already been parallelized previously. Nevertheless, parallelization of all stages is necessary in order to achieve high efficiency.  

For the first time, we present a decoder pipeline which is fully implemented on the GPU. By using this implementation, we demonstrate that the presented \textit{overflow pattern} is suitable for decompressing JPEG-encoded data on the GPU at higher throughput rates in comparison to the nvJPEG hardware accelerated decoder implementation by NVIDIA. In addition, we provide deeper insights on factors impacting performance and present experimental results on datasets used in practice. In particular, our method shifts the bottleneck from Huffman decoding to other stages of the pipeline. Our performance evaluation using datasets with images of various sizes and image qualities shows that our implementation achieves speedups of up to 51 (34) over CPU-based libjpeg-turbo and 8.0 (5.7) over nvJPEG on an A100 (V100) GPU, respectively. Furthermore, we outperform the hardware-accelerated version of nvJPEG on an A100 by a factor of up to 3.4. The contributions of our paper are threefold:
\begin{itemize}
    \item Parallelization of the complete JPEG decoder pipeline
    \item Discussion of implementation details of the proposed algorithm
    \item Performance evaluation and comparison against state-of-the-art decoder libraries on modern hardware
\end{itemize}

The rest of this paper is organized as follows: Section \ref{sec:related} reviews related work. Section \ref{sec:jpeg} describes the process of encoding and decoding JPEG files and elaborates further on the JPEG specification.  Our algorithm for parallel decoding is presented in Section \ref{sec:parallel}. Experimental results are presented in Section \ref{sec:experiments}. Section \ref{sec:conclusion} concludes and proposes opportunities for future research.

\section{Related Work} \label{sec:related}
In the past, there have been multiple efforts addressing the parallelization of JPEG decoding.

Yan et al.~\cite{yan2013cuda} accelerated JPEG decoding on GPUs by parallelizing the IDCT step using CUDA. Sodsong et al.~\cite{sodsong} solved the problem of bitstream decoding by letting the host system determine the positions of individual syntax elements (bit sequences). Using this information, the actual decoding can be performed in parallel on the GPU. However, the requirement of processing the compressed data entirely on the host side can still represent a bottleneck. Wang et al. \cite{wang2020accelerating} proposed two prescan methods to eliminate the unnecessarily repetitive scanning of the JPEG entropy block boundaries on the CPU but only achieve a speedup of 1.5 over not hardware-accelerated nvJPEG. GPUJPEG \cite{holub2013gpu} takes advantage  of restart markers that allow for fast parallel decoding. Unfortunately, these markers are not used in JPEG-encoded images by most libraries and cameras, which limits the use cases of this approach.

In recent years, a lot of research has been accomplished regarding GPU-based compression and decompression of Huffman encoded data in general~\cite{sitaridi}~\cite{patel}~\cite{tian}.

FPGAs and processor arrays have also been used as alternative architectures for JPEG \cite{cheng2019dlbooster} and Huffman decoding \cite{sarangi2021canonical}. Different from GPU-based solutions these approaches focus on energy efficiency by designing
optimized processing pipelines or bit-parallel architectures in hardware.

Klein and Wiseman~\cite{klein_wiseman} constructed a parallel algorithm for multi-core processors capable of decoding data that has been compressed using Huffman's original method. It relies on the so-called \textit{self-synchronizing property}, which many Huffman codes possess. They also showed that this technique is applicable to JPEG files. Recently, this method was used to create a fully GPU-based Huffman decoder~\cite{weissenberger}.

In Section 4, we introduce a novel approach to JPEG bitstream decoding based on the work of Klein and Wiseman. We will also show how this method can be used to create a fully parallel GPU-based JPEG decoder which can exploit the large number of cores available on modern GPUs. This will build upon the techniques presented in \cite{weissenberger} and further extend them. We will provide a specification of the resulting algorithm and evaluate its implementation in experiments.

\section{JPEG Encoding and Decoding}\label{sec:jpeg}

\begin{figure*}
\centerline{\includegraphics[width=\textwidth]{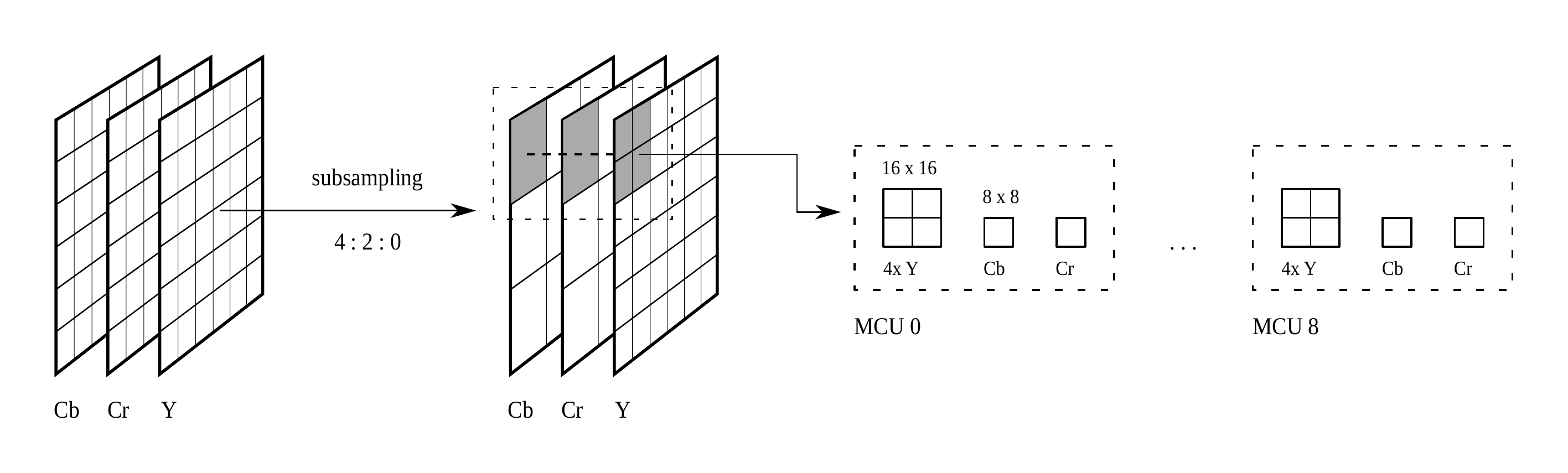}}
\caption{Example for 4:2:0 chroma subsampling (chroma resolution is quartered) and subsequent grouping of data units into MCUs.}
\label{figure_subsampling}
\end{figure*}

The JPEG encoding process consists of the following steps:
\begin{enumerate}
\item Color space conversion
\item Chroma subsampling
\item Decomposition of channel planes into blocks of 8x8 pixels.
\item Forward Discrete Cosine Transform (DCT)
\item Quantization
\item Differential DC encoding
\item Zig-zag transform
\item Run-length encoding
\item Huffman encoding
\end{enumerate}

In the first step, the source image is converted from RGB (red, green and blue pixel format) to the YCbCr colorspace. This leads to decorrelation of the brightness (luminance, Y) and color components (chrominance, Cb and Cr).

In the following, the chrominance part will be compressed at higher loss than the luminance part. Initially, \textit{subsampling} is applied to the chroma planes, i.e., their resolution is reduced by discarding rows or columns of pixels. The subsampling modes most commonly used by software libraries and cameras are referred to as \textit{4:2:2} (chroma resolution halved horizontally) and \textit{4:2:0} (chroma resolution halved horizontally and vertically). The detailed requirements of subsampling and applicable subsampling modes are specified in \cite{cipa} and \cite{jfif}. The luminance component is excluded from subsampling and thus remains uncompressed during this first stage.

Each component is then divided into blocks of 8x8 pixels, referred to as \textit{data units}. In the following, each data unit will be processed independently. Depending on the respective chroma subsampling mode, a varying number of locally corresponding data units \textit{from different components} is grouped together. Such a group is referred to as a \textit{Minimum Coded Unit (\textbf{MCU})}. In case of the common horizontal 4:2:2 subsampling, MCUs have a total size of 16x8 pixels, and consist of two horizontally adjacent luminance data units, as well as a single Cb and Cr data unit. Thereby, each chrominance sample represents an average value of two neighboring pixels. In case of 4:2:0 subsampling, MCUs have a total size of 16x16 pixels, and consist of four luminance data units (organized in two rows and two columns) and again a single Cb and Cr data unit. Since the chrominance resolution is quartered, each one of the chrominance samples represents an average value of a block of 2x2 pixels. Figure \ref{figure_subsampling} illustrates the process of 4:2:0 subsampling applied to an image of size 48x48 pixels, as well as grouping the resulting data units into nine MCUs. If no chroma subsampling is used (also known as 4:4:4 subsampling), the chrominance planes are stored at full resolution. In this case, each MCU has a resolution of 8x8 pixels and is composed of a single luminance data unit, and two single chrominance units. For image dimensions not representing multiples of the MCU size, any incomplete MCU must be augmented with generated pixels. The JPEG standard recommends to insert repetitions of the rightmost column (or bottom row) of pixels from the image to complete MCUs. Those excess pixels will be discarded (or ignored) after decoding.

Following chroma subsampling, each data unit of 8x8 pixels is transformed to the frequency domain independently, using a Type II discrete cosine transform (DCT). The output of this process is a matrix of 64 transform coefficients. The coefficient in the upper left corner represents the direct current component of the block, and thus is referred to as \textit{DC coefficient}, while all others are referred to as \textit{AC coefficients}.  In the following \textit{quantization} step, this representation allows for the reduction of the magnitude of single frequencies, or even to discard them entirely.
In order to quantize a block, each transform coefficient $t_{ij}$ is divided by a corresponding value $Q_{ij}$ from a quantization matrix $Q$. The resulting value $q_i$ is referred to as \textit{quantized coefficient}, $q_i = \left[\frac{t_{ij}}{Q_{ij}}\right]$.
The matrix $Q$ is supplied by the encoder and will be embedded in the header of the compressed file. JPEG allows the use of up to two different quantization matrices. Usually, the first matrix is applied to the luminance component, the second matrix to the chrominance components.

In the quantization step, image quality will be reduced as frequencies are discarded. However, by choosing appropriate quantization tables, relevant frequencies will remain mostly unaltered while less relevant frequencies will be removed. As the JPEG compression scheme has been optimized for compressing photographic images, quantization tables should usually reduce high frequency coefficients, and preserve low frequency coefficients. When a suitable $Q$ is used, the decoded image will appear lossless to the human visual system, while its size will be significantly reduced.

Subsequent to quantization, the DC coefficient and the 63 AC coefficients will be processed separately for each data unit. The current DC coefficient $DC_{i}$ is predicted and replaced by $P$, the difference between $DC_{i}$ and $DC_{i-1}$, i.e., $P = DC_{i} - DC_{i-1}.$

\begin{figure}
\centerline{\includegraphics[scale=0.25035, trim= 0 40 0 50, clip=true]{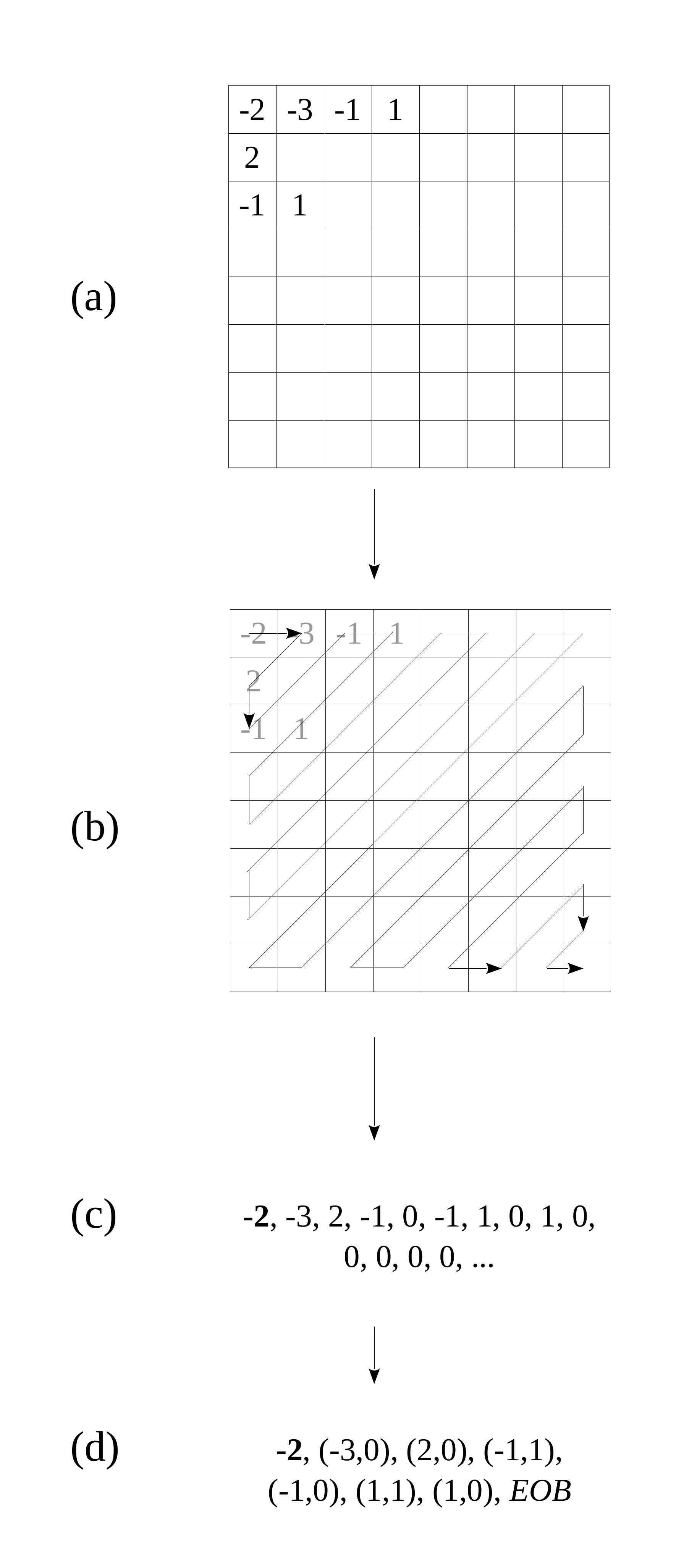}}
\caption{Zig-zag transform and run-length encoding of a quantized block of 8x8 coefficients}
\label{figure_runlength}
\end{figure}

$DC_0$ is not replaced and will be used to initialize the prefix sum during decoding to reverse the process. The AC coefficients, in contrast, will be scanned in a so-called "zig-zag" order, as depicted in Figure \ref{figure_runlength}. Figure \ref{figure_runlength} (a) illustrates a typical example for a data unit from a photographic image. Zero-valued coefficients have been replaced with empty cells for improved readability. Figure \ref{figure_runlength} (b) shows the process of zig-zag reordering the AC coefficients. The output sequence is shown in (c). As the DCT typically concentrates a major part of the signal energy in the upper left corner, longer runs of zero coefficients appear in the remaining places. This redundancy is removed immediately afterwards using run-length encoding. Each AC coefficient $AC_i$ is replaced by a tuple $(AC_i, \text{ZRL}_i)$, where $\text{ZRL}_i$ represents the number of zeros preceding the $i$-th coefficient. The size of $\text{ZRL}_i$ is limited to $15$. In case all remaining coefficients in the block ($i\dots 63$) represent a zero run, encoding is finalized by a special symbol (\textbf{EOB}, \textit{end of block}). Another special symbol (\textbf{ZRL}, \textit{zero run length}) is used to indicate a run of $16$ zeros. For runs longer than $16$, multiple $\text{ZRL}$ symbols can be concatenated. The generated sequence for the example is shown in Figure \ref{figure_runlength} (d).

\begin{table*}
    \caption{Basic example of bitstream synchronization in JPEG}
    \begin{center}
\begin{tabular}{@{}ccccccccc@{}}
\toprule
\textbf{(2,0) -2} & \multicolumn{2}{c}{(2,0) -3} & (2,0) 2 & (1,0) -1 & (1,1) -1 & (1,0) 1 & (1,1) 1 & EOB \\
\textbf{\underline{10}01} & \multicolumn{2}{c}{\underline{100}00} & \underline{100}10 & \underline{01}0 & \underline{1011}0 & \underline{01}1 & \underline{1011}1 & \underline{00} \\ \midrule
\underline{10}01 & \textbf{\underline{10}00} & \underline{01}0 & \underline{01}0 & \underline{01}0 & \underline{1011}0 & \underline{01}1 & \underline{1011}1 & \underline{00} \\
 &  \multicolumn{1}{l}{$\uparrow$} &  &  & \multicolumn{1}{l}{$\circ$} &  &  &  &  \\
 & \textbf{(2,0) -3} & (1,0) -1 & (1,0) -1 & (1,0) -1 & (1,1) -1 & (1,0) 1 & (1,1) 1 & EOB \\ \bottomrule
\end{tabular}
    \end{center}
    \label{tab:parallel}
\end{table*}

Afterwards, the sequences of tuples in each block are Huffman encoded~\cite{huffman} to produce the final bitstream. In the case of baseline JPEG, two different Huffman trees are created for the luminance component, and two additional trees for the chrominance components. For each component, the DC coefficients are encoded by the first tree, while the AC values are encoded by the second tree. The resulting bitstream consists of an alternating sequence of Huffman codewords and signed ones' complement integers, each of which represents one of the transform coefficients. The preceding codeword encodes a tuple of the form $(l, r)$, where $l$ indicates the length of the following binary representation, and $r$ represents the count of zeros preceding the encoded coefficient. For example, if the codeword $100$ signified the information
$l = 2, r = 0,$
then the first AC coefficient in Figure \ref{figure_runlength}, $-3$, would be encoded as $10000$.
The EOB symbol is represented by $(l = 0, r = 0)$, while ZRL is represented by ($l = 0, r = 15$). No run-length value is encoded for DC coefficients. Table \ref{tab:parallel} lists a complete encoding for the example in Figure \ref{figure_runlength}.

The JPEG decoding process reverses the nine steps mentioned above loss-free with the exception of rounding errors and the quantization step, which is lossy.

\section{Parallel Decoding}\label{sec:parallel}

When dividing the JPEG-decoded bitstream of each image into multiple segments for distribution to different processors, data dependencies between segments may occur because of the following reasons:

\begin{itemize}
\item A segment boundary was placed within a Huffman codeword
\item A segment boundary was placed within a coefficients' binary representation
\item The correct Huffman table for the current data is unknown
\item The current index in the zig-zag sequence is unknown
\end{itemize}

Klein and Wiseman suggested an algorithm~\cite{klein_wiseman} to resolve these dependencies. Their approach relies on the observation that certain Huffman codes tend to \textit{resynchronize} after transmission errors. When an error occurs, a decoder produces erroneous symbols. However, if the respective Huffman code is \textit{self-synchronizing}~\cite{ferguson}, correct symbols will be produced after a certain number of bits have been processed. Fortunately, Huffman codes produced by JPEG encoders possess this property.

Consider the example listed in Table \ref{tab:parallel} for better illustration. The upper part contains the sequence of coefficients from Figure \ref{figure_runlength}. The first row lists the syntax elements, while the second row shows the corresponding binary encoding of the same element. The DC coefficient is set in boldface, Huffman codes are underlined.

The lower part of the table shows the result of decoding when starting at the fifth bit within the input (marked with the arrow). In this case, the process starts at in between two syntax elements. However, the wrong table is used as the decoder expects a DC coefficient. After the first value has been decoded, the AC table will be used until the end of the block. Subsequently, two more incorrect symbols will be decoded. However, the last incorrectly interpreted binary string forms a suffix of the correct string at the same position. This implies that the next symbol, and as a consequence, the rest of the block, will be decoded correctly. The point at which the two processes became synchronous is referred to as a \textit{synchronization point} (indicated by the circle).

The self-synchronizing property is exploited in~\cite{klein_wiseman} to enable parallel decoding. Following the division of the input into equally sized segments, a processor is assigned to each segment to decode the contained data. After a given processor has finished decoding, it will overflow to the adjacent segment  and replace symbols which were erroneously decoded by the succeeding processor, until it detects a synchronization point. If no synchronization point was detected, the processor will continue to overflow to the next segments, until the end of the input is reached.

The decoder presented in this paper can be considered as a function mapping a multiplicity of encoded JPEG bitstreams (i.e. a batch of JPEG files) to an array of pixels. Each of those arrays represents a separate color channel and contains the sequence of decoded images for the respective channel. The decoder consists of multiple components: Combined Huffman- and run-length decoding, differential- and zig-zag decoding, dequantization, inverse DCT.

\subsection{Combined Huffman- and Run-Length Decoding}

\input{alg_decoding}

\input{alg_single}

\input{alg_sync}

Algorithm \ref{alg:write_output} lists all of the steps required for Huffman- and run-length decoding. Initially, after parsing the batch and extracting all necessary Huffman tables, each individual bitstream is divided into multiple $32\cdot s $-bit sized chunks. In the following, those chunks will be referred to as \textit{subsequences}. From the resulting $N$ subsequences, $b$ adjacent subsequences form a \textit{sequence}, for a total of $B$ sequences (Lines 2-4). \texttt{s\_info}, allocated in Line 5, represents a globally accessible buffer of size $N$. Its $i$-th entry will contain all required information about the last sychronization point in subsequence~$i$.

Each sequence will in the following be decoded by a thread block of size $b$, with each thread decoding an individual subsequence. Algorithm \ref{alg:subsequence} lists the process of decoding a single subsequence. Lines 11 - 27 represent the mechanism to decode JPEG syntax elements. The function \texttt{decode\_next\_symbol()} in Line 13 extracts coefficients from the bitstream while switching between DC and AC Huffman tables. It returns the number of processed bits, the decoded coefficient (provided the code was not \textbf{EOB} or \textbf{ZRL}) and the run length of zeroes which the coefficient is followed by. If the \textit{write} flag is set, the coefficient is written to the output buffer (Lines 14 - 16, see below). In case the symbol equals \textbf{ZRL}, $15$ will be returned for run\_length. If the symbol equals \textbf{EOB}, $63 - z$ will be returned, with $z$ being the current index in the zig-zag sequence. After the symbol has been decoded, $p$ bits are skipped (Line 18) and the number of decoded symbols as well as the zig-zag index are updated (Lines 19 - 20). The zig-zag index is set to zero if the data unit is complete (Line 21 - 23). Depending on chroma subsampling, the color component $c \in \{Y,Cb,Cr\}$ is changed after the completion of a data unit, if necessary (Lines 24 - 26). For example, if \textit{4:2:0} subsampling is used in the image, $c$ would change from $Y$ to $Cb$ after 4 data units, from $Cb$ to $Cr$ after another data unit and back to $Y$ after another data unit. Considering 4:4:4 subsampling,  $c$ would change after every decoded data unit.

The process that is described in the following is referred to as ``intra-sequence synchronization''. After decoding of subsequence $i$ is complete, the corresponding thread will store the location of the last detected codeword ($p$) in $i$, as well as the current position in the zig-zag sequence ($z$) and the current color component ($c$) at the $i$-th field of \texttt{s\_info}$[i] \longleftarrow (p, z, c)$ (Algorithm \ref{alg:decoder}, Line 10). Next, the thread waits for the data of subsequence $i+1$ to be available in \texttt{s\_info} (marked by the \textit{\_\_syncthreads()} function in Line 11), and continues decoding the following subsequence, $i + 1$ afterwards (Line 13 - 22). This is referred to as an \textit{overflow}. If there is a synchronization point in subsequence $i + 1$, the decoding process will resynchronize, and the values $p$, $z$ and $c$ will each be identical to those stored in \texttt{s\_info}$[i+1]$. If thread $i$ detects this (Lines 16 - 18), it indicates synchronization by setting the flag in Line 17. As subsequence $i+2$ would be decoded correctly, it pauses and waits for the other threads in the same thread block to finish (Line 21).

If no synchronization points exist in subsequence $i+1$, thread $i$ stores the updated values $(p,n,z,c)$ (along with the number of decoded symbols $n$) in \texttt{s\_info}$[i+1]$ (Line 19). It repeats the process by overflowing to the following subsequences until it either detects synchronization or reaches the end of the last subsequence within the sequence (loop at Line 13).

Next, ``inter-sequence synchronization" follows to establish synchronization between sequences. For this purpose, one thread is assigned to each sequence, except for the last (Line 25). Afterwards, thread $i$ overflows out of the last subsequence of sequence $i$ into the first subsequence of sequence $i+1$. Next, the thread continues overflowing into the following subsequences until, again, it either detects synchronization or reaches the end of the last subsequence in the sequence (loop at Line 30). To indicate that synchronization has been achieved, the thread sets a flag in a globally accessible array, \texttt{sequence\_synced} (Line 35).

The host system repeats the inter-sequence synchronization phase until all flags in \texttt{sequence\_synced} are set (loop at line 25). Afterwards, correct parallel decoding is possible using the information from \texttt{s\_info}.

Before any final output is written, an exclusive prefix sum is computed on the fields storing the symbol count ($n$) of \texttt{s\_info} (Algorithm \ref{alg:write_output}, Lines 7 - 8). Afterwards, \texttt{s\_info}$[i].n$ will contain the correct index in the output buffer for the decoded content of subsequence $i$ to start.

Finally, the decoded output is written, first by again assigning a thread to each subsequence. Then, thread $i$ consults \texttt{s\_info}$[i]$ to retrieve the location of the first codeword in this subsequence, set the correct component and correct index in the zig-zag sequence, and write the output to the correct location in the output buffer (Lines 9 - 15). \vspace{-4pt}

\subsection{Difference Decoding}
\vspace{-1pt}
In order to reverse DC difference coding, a prefix sum must be computed on the DC coefficients of each component for all images in the batch (Algorithm \ref{alg:write_output}, Lines 16 - 18).

For our implementation evaluated in Section \ref{sec:experiments}, this is achieved using a parallel scan method from the CUDA \textit{thrust} library, applied to each component separately, with a custom operator. This allows for the correct indices to be included in the scan, as well as to reinitialize the sum at the end of each image.

\subsection{Zig-zag Decoding, IDCT and Dequantization}

After difference decoding is completed, zig-zag decoding is applied to every data unit (block of $8\times 8$ pixels) in every image in the batch. This merely represents a permutation that is applied to each data unit independently, and therefore is trivially parallel. Zig-zag decoding is also available on GPUs as part of the NVIDIA Performance Primitives Library (NPP).

Next, the data units are individually dequantized. This is done by multiplying each quantized coefficient $q_{ij}$ from the data unit by its corresponding entry $Q_{ij}$ from the appropriate quantization matrix in the file header.

The Inverse Discrete Cosine Transform (IDCT) is also applied to each data unit independently (Algorithm \ref{alg:write_output}, Lines 19 - 21). Efficient implementation of the and IDCT on GPUs has been subject of multiple research works, including \cite{Obukhov}. 

We implemented the inverse zig-zag and cosine transform, as well as dequantization in a single kernel. Each thread is thereby configured to process a single $8\times 8$ data unit.

The IDCT marks the last step in the JPEG decoding pipeline (as long as no subsampling or colorspace conversion was applied to the image). However, the unchanged output consists of a series of MCUs, each containing multiple data units from multiple components. In order to retrieve complete components, contiguous values can be extracted from the MCUs and then be written to the appropriate buffer in the correct order. In our implementation, three separate kernels are used to copy the values for different components. However, efficiency is limited as memory accesses are not aligned.

\section{Experimental Results}\label{sec:experiments}

We have implemented our JPEG decoder (in the following discussion referred to as 'jgu') using CUDA C/C++ 11.2 and tested it on the following systems:
\begin{itemize}
    \item System 1 (V100): Intel Xeon Gold 6238 CPU @ 2.10 GHz, 3.7 GHz (boost), 22 cores, 192 GB RAM, NVIDIA Quadro GV100 equipped with 32GB of VRAM.
    \item System 2 (A100): NVIDIA DGX A100, AMD Epic Rome 7742 CPU @ 2.25 GHz, 3.4 GHz (boost),  128 cores, 1 TB RAM, 8x NVIDIA A100 GPUs, each equipped with 40 GB of VRAM, special decoder hardware for JPEG files.
\end{itemize}

Time measurements are accomplished with CUDA event
system timers. In all experiments, we assume that the encoded image data resides on the host system, and decoded image data resides on the GPU. Time spent on transferring compressed image bitstreams to the device, as well as on GPU-sided malloc- and free operations for auxiliary memory, is included in the timings.

\subsection{Libraries}

In order to evaluate the performance of our approach, we compare it to two state-of-the-art JPEG decoders: libjpeg-turbo~\cite{libjpegturbo}, version \textit{2.0.6} and nvJPEG~\cite{nvjpeg}, part of the CUDA toolkit, version \textit{11.3.0}. The latter represents NVIDIA’s GPU-based decoder and is included in the NVIDIA Data Loading Library (DALI). It is frequently used to improve throughput in Deep Learning applications and therefore is best suited to processing of large batches containing images of small or medium size. 

libjpeg-turbo is an open-source, CPU-based, single-threaded decoder featuring specialized, high performance Huffman decoding routines. It accelerates the calculation of the IDCT by the use of vector registers.

\subsection{Datasets}

The real-world datasets used for the following experiments, as well as their specific properties, are listed in Tables \ref{tab:various} and \ref{tab:quality}. Each dataset consists of a single batch of images. A batch represents a sequence of photographic images extracted from video footage at different resolutions and qualities. \textbf{stata} and \textbf{newyork} (Table \ref{tab:various}) contain footage from hand cameras and have been chosen as prototypical examples for datasets used in deep learning applications, e.g. classification of objects.

\textbf{tos\_4k} and \textbf{tos\_1440p} represent the same sequence from the open movie \lq Tears of Steel\rq (2013) at two different resolutions. The images have been extracted from the video using the tool \textit{ffmpeg}. \textbf{tos\_8}, \textbf{tos\_14} and \textbf{tos\_20} represent the same sequence at reduced image quality (i.e. scaling entries of quantization matrices by a different factor). The numbers represent the \textit{ffmpeg} quality parameter (-qscale:v) used for encoding. All of the other datasets (including \textbf{tos\_4k}) have been encoded at maximum quality (-qscale:v 2).

\begin{table*}
\caption{Properties of datasets used in the experiments featuring varying resolutions and batch sizes.}
\begin{tabular}{lllllll}
           & batch size & subsequence size & resolution X & resolution Y & batch size (compressed, MB) & average image size (KB) \\
newyork    & 500        & 1024             & 1920         & 1080         & 134.1                       & 250                     \\
stata      & 2400       & 1024             & 720          & 480          & 550                         & 230                     \\
tos\_1440p & 200        & 1024             & 2560         & 1440         & 84                         & 420                     \\
tos\_4k    & 200        & 1024             & 3840         & 2160         & 100                            & 550 
\end{tabular}
\label{tab:various}
\end{table*}

\begin{table*}
\caption{Properties of datasets used in the experiments featuring varying image quality.}
\begin{tabular}{lllllll}
        & batch size & subsequence size & resolution X & resolution Y & batch size (compressed, MB) & average image size (KB) \\
tos\_8  & 200        & 128              & 2560         & 1440         & 28.5                        & 180                     \\
tos\_14 & 200        & 1024             & 2560         & 1440         & 21.9                        & 130                     \\
tos\_20 & 200        & 1024             & 2560         & 1440         & 18.9                        & 110                    
\end{tabular}
\label{tab:quality}
\end{table*}

\subsection{Performance Breakdown}

\begin{figure}
    \centering
    \includegraphics[scale=0.52, trim = 5 5 5 5, clip=true]{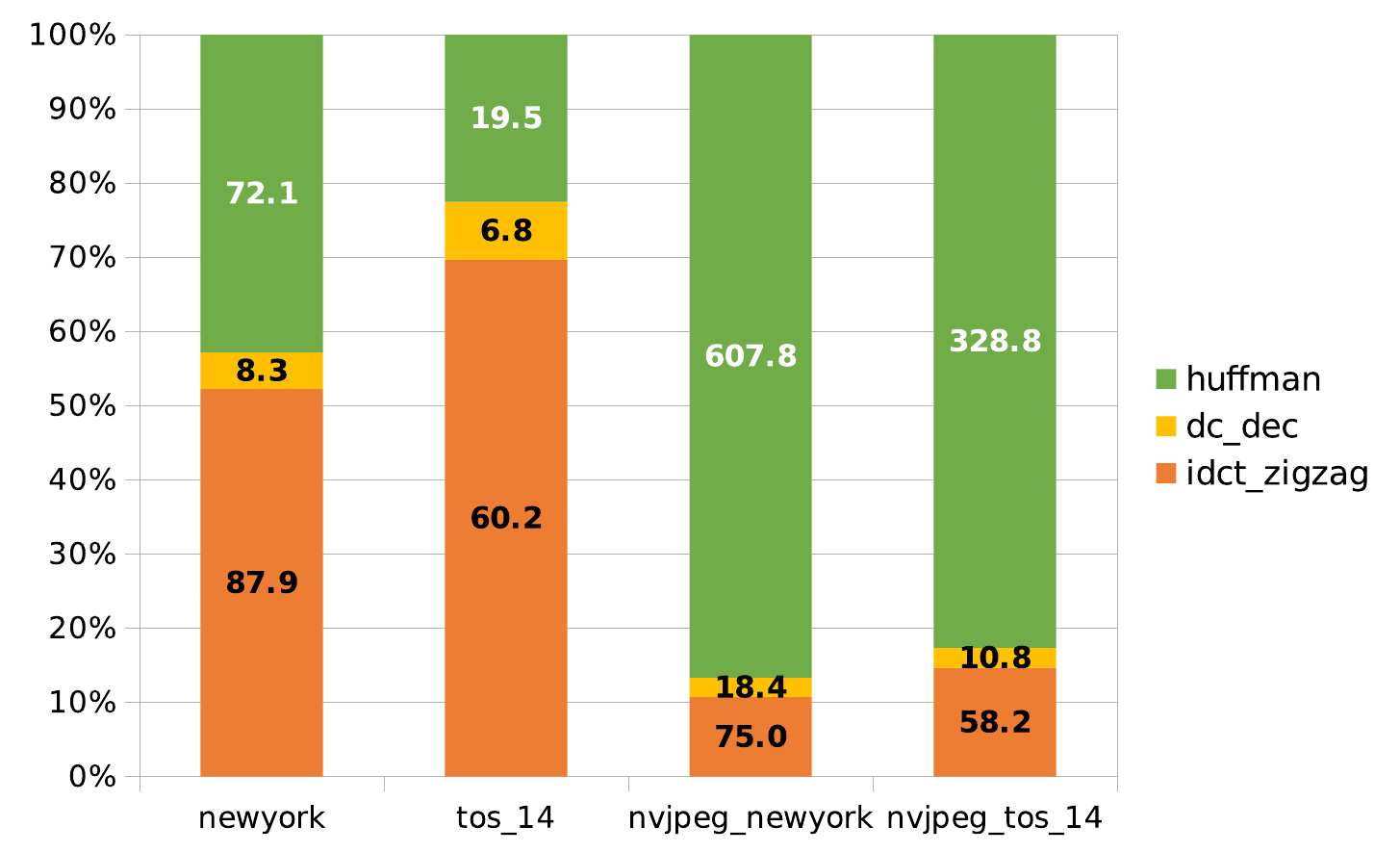}
    \caption{Percentual breakdown of runtimes of our decoder and nvJPEG for two datasets. The values in the bars indicate the absolute runtime of the respective task in milliseconds (ms).}
    \label{fig:breakdown}
\end{figure}

We have analyzed the runtime behavior of our JPEG decoder on a V100 GPU in detail. 
Figure \ref{fig:breakdown} illustrates the shares in the total runtime of the components of our JPEG decoder pipeline for two representative datasets: newyork and tos\_14. In this scenario, the datasets were decoded using subsequence sizes of 1024 and 128 respectively. Since tos\_14, in comparison to newyork, is compressed at higher loss, it features fewer syntax elements and thereby fewer and shorter Huffman codes. Thus, decoding can synchronize sooner, which allows for smaller subsequence sizes and more parallel threads. In addition, the number of write operations per thread is reduced, resulting in faster Huffman decoding. In contrast, the newyork dataset is of high quality, so the contained bitstreams consist of numerous, long Huffman codes, synchronization is therefore prolonged.

In Figure \ref{fig:breakdown}, the bars at the left-hand side represent the results for our own implementation, while the bars on the right-hand side represent the results of the nvJPEG library on the same datasets, as indicated by the prefix \textit{nvjpeg\_}. ‘huffman' represents the percentage of total runtime spent on Huffman decoding. While for newyork the task consumes $42.8\%$ ($72$ ms), for the tos\_14 dataset, merely $23.1\%$ ($19.46$ ms) of the time are spent on Huffman decoding by our implementation. nvJPEG respectively spends $86.7\%$ ($607.8$ ms) and $82.7\%$ ($328.81$ ms) on the same datasets. This indicates that the Huffman decoding stage is handled efficiently by our implementation, while representing a bottleneck for nvJPEG. dc\_dec represents the time used to adjust the DC coefficients, idct\_zigzag represents the part of the runtime spent on computing the IDCT and performing zig-zag decoding as well as dequantization.

Our Huffman decoding implementation can be further broken down into the three individual parts: (i) intra sequence synchronization (Algorithm \ref{alg:decoder}, Lines 1 - 24), (ii) inter sequence synchronization (Algorithm \ref{alg:decoder}, Lines 25 - 43), and (iii) writing decoded data (Algorithm \ref{alg:write_output}, Lines 9 - 15). For the newyork dataset, $39.57\%$ ($28.5$ ms) of the Huffman runtime were spent on intra sequence synchronization, $14.44\%$ ($10.4$ ms) on inter sequence synchronization, $45.99\%$ ($33.1$ ms) on writing decoded data. For tos\_14, $25.92\%$ ($5.2$ ms) were spent on intra sequence synchronization, $3.23\%$ ($0.6$ ms) on inter sequence synchronization and $70.85\%$ ($14.2$ ms) on writing decoded data.

\subsection{Performance Comparison}

On the charts discussed in this section, ‘nvjpeg\_V100’ (‘nvjpeg\_A100’) denotes the nvJPEG GPU-based decoder executed on a V100 (A100) GPU, while ‘nvjpeg\_hw’ denotes nvJPEG executed using the special decoder hardware that is part of the A100 GPU. ‘libjpegturbo’ denotes the CPU-based single-core open source JPEG decoder executed on an AMD Epic Rome CPU of System 2.

\begin{figure}
    \centering
    \includegraphics[scale=0.52, trim= 0 8 0 5, clip=true]{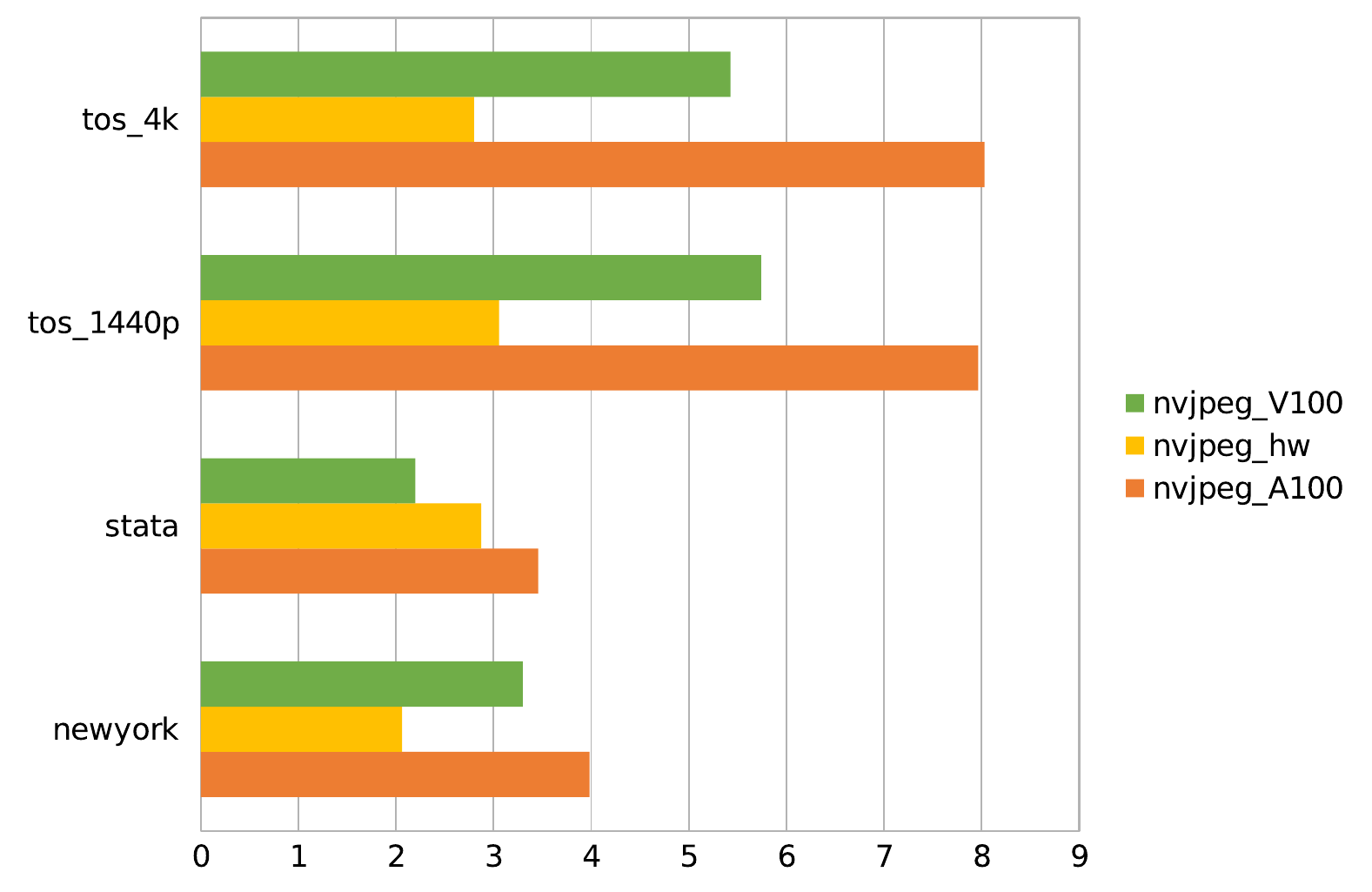}
    \caption{Speedup over nvJPEG for various datasets}
    \label{fig:speedup_various_nvjpeg}
\end{figure}

\begin{figure}
    \centering
    \includegraphics[scale=0.52, trim= 0 8 0 5, clip=true]{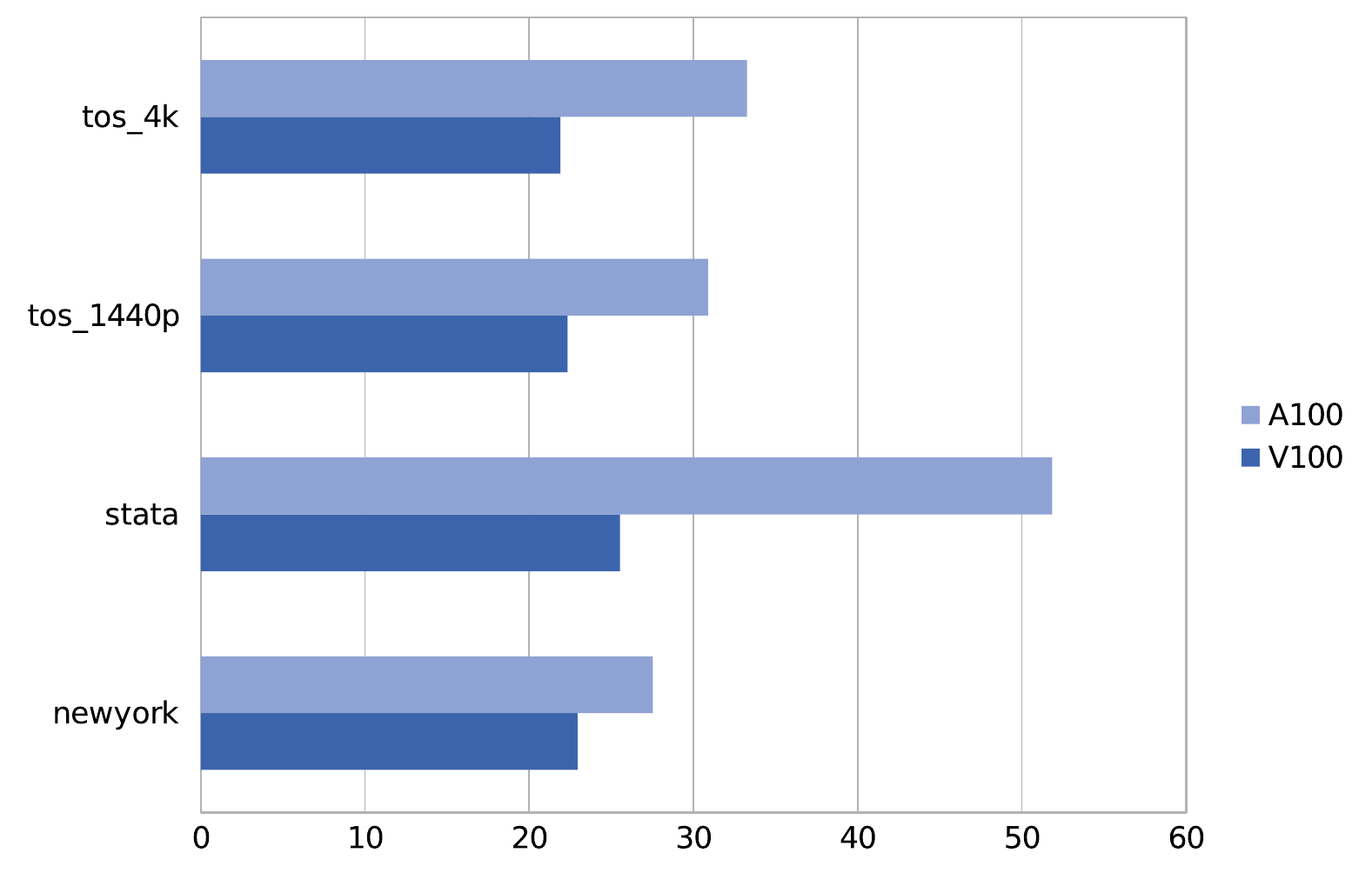}
    \caption{Speedup over libjpegturbo for various datasets}
    \label{fig:speedup_various_libjpegturbo}
\end{figure}

Figures \ref{fig:speedup_various_nvjpeg} and \ref{fig:speedup_various_libjpegturbo} show the speedups achieved by ‘jgu’ over nvJPEG and libjpegturbo for various datasets. In a first scenario, ‘jgu’ and ‘nvjpeg\_V100’ are compared using System~1. In a second scenario, ‘jgu’, ‘nvjpeg\_A100’ and ‘nvjpeg\_hw’ are compared using System~2. All tests only use a single GPU.

While the newyork dataset decodes more than 3 times faster on the V100, and almost 4 times faster on the A100 compared to nvJPEG, speedups of 2.2 and 3.4 are observed for the stata dataset, respectively. The tos datasets were decoded with a speedup between 5 and 6 on the V100 and a speedup of 8 on the A100. As stata contains considerably more images than there is in any other dataset, ‘jgu’ delivers the highest speedup over libjpegturbo on this dataset (over 50), as the GPU is, unlike the host CPU, able to decode the majority of images in the batch in parallel. The achieved speedups using both tos datasets are similar since the images contained differ in resolution, but otherwise have similar features.

\begin{figure}
    \centering
    \includegraphics[scale=0.52, trim= 0 8 0 5, clip=true]{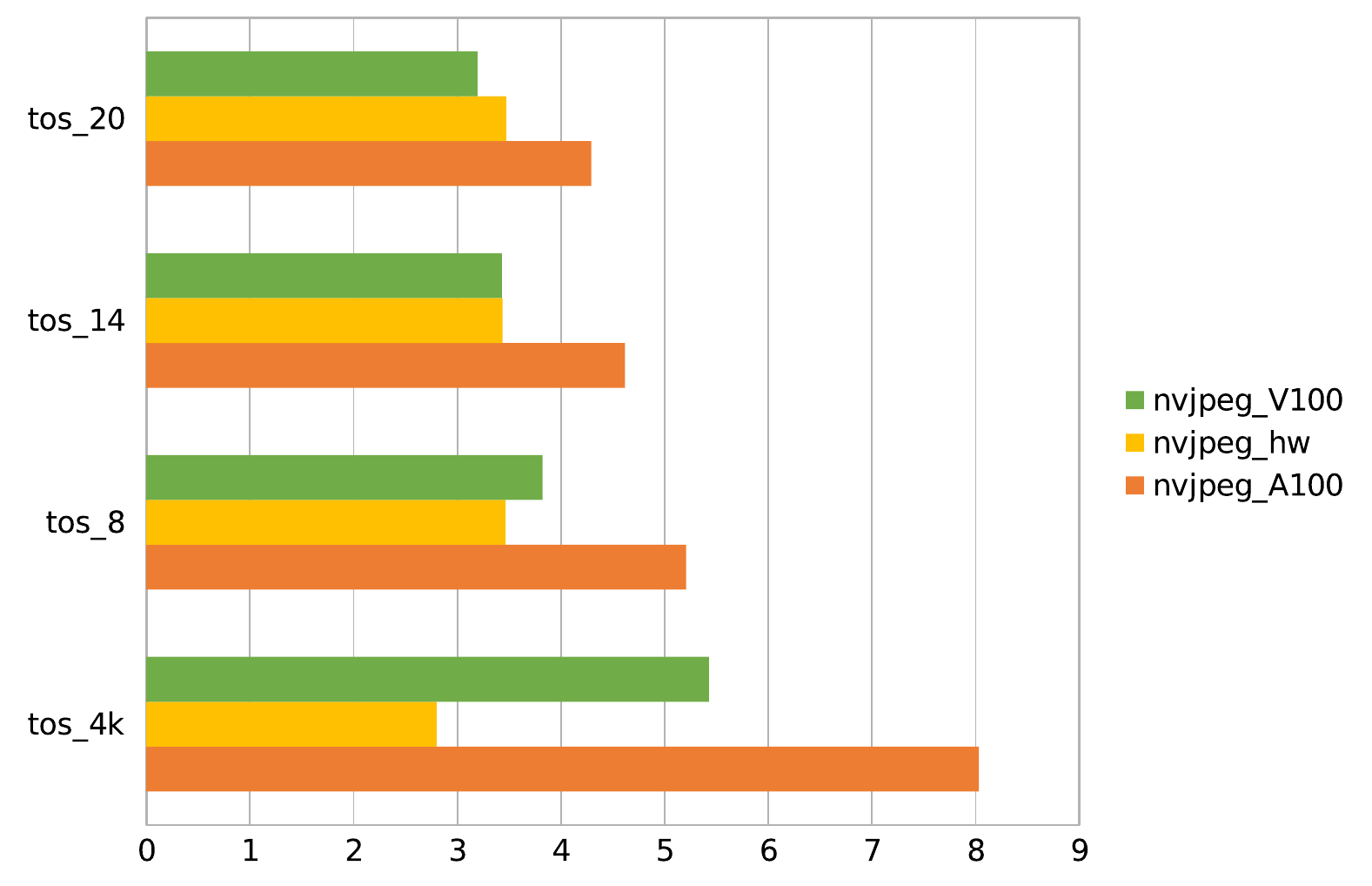}
    \caption{Speedup over nvJPEG for various image qualities}
    \label{fig:speedup_quality_nvjpeg}
\end{figure}

\begin{figure}
    \centering
    \includegraphics[scale=0.52, trim= 0 8 0 5, clip=true]{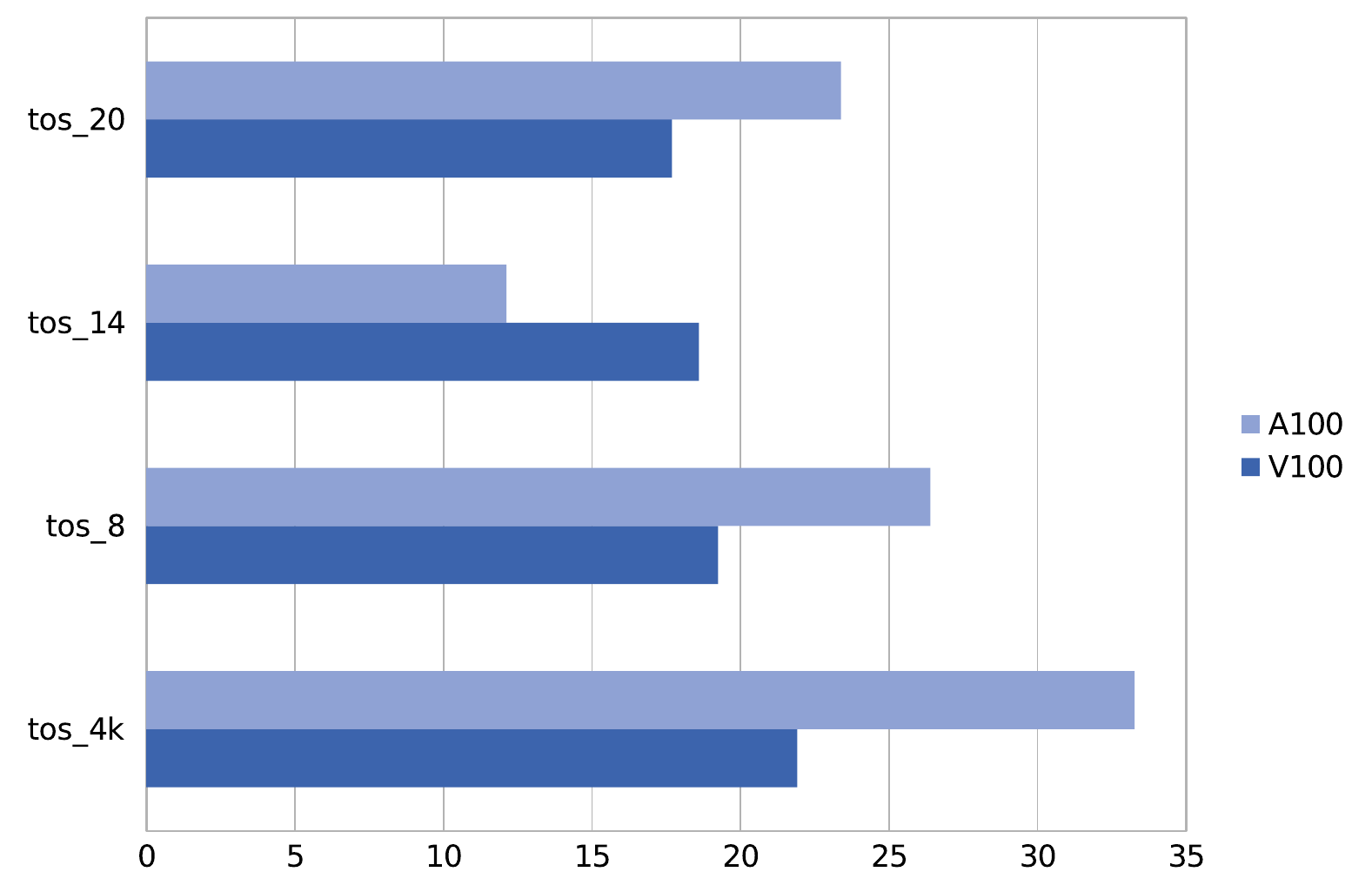}
    \caption{Speedup over libjpegturbo for various image qualities}
    \label{fig:speedup_quality_libjpegturbo}
\end{figure}

Figures \ref{fig:speedup_quality_nvjpeg} and \ref{fig:speedup_quality_libjpegturbo} show the speedups of ‘jgu’ over nvJPEG and libjpegturbo for four datasets, tos\_$x$ (with $x \in \{8, 14, 20\}$) and tos\_4k, of varying image quality. $x$ equals the ffmpeg quality setting used to encode the batch. Higher values of $x$ lower the image quality but increase compression rates. Furthermore, tos\_4k represents the maximum image quality for this dataset. It can be seen that the achieved speedups over the non-hardware accelerated nvJPEG versions in Figure \ref{fig:speedup_quality_nvjpeg} generally decrease for lower quality images. In these cases, shorter bitstreams yield quicker decoding for nvJPEG, while the number of images remains constant across different datasets.

\begin{figure}
    \centering
    \includegraphics[scale=0.52, trim= 0 5 0 5, clip=true]{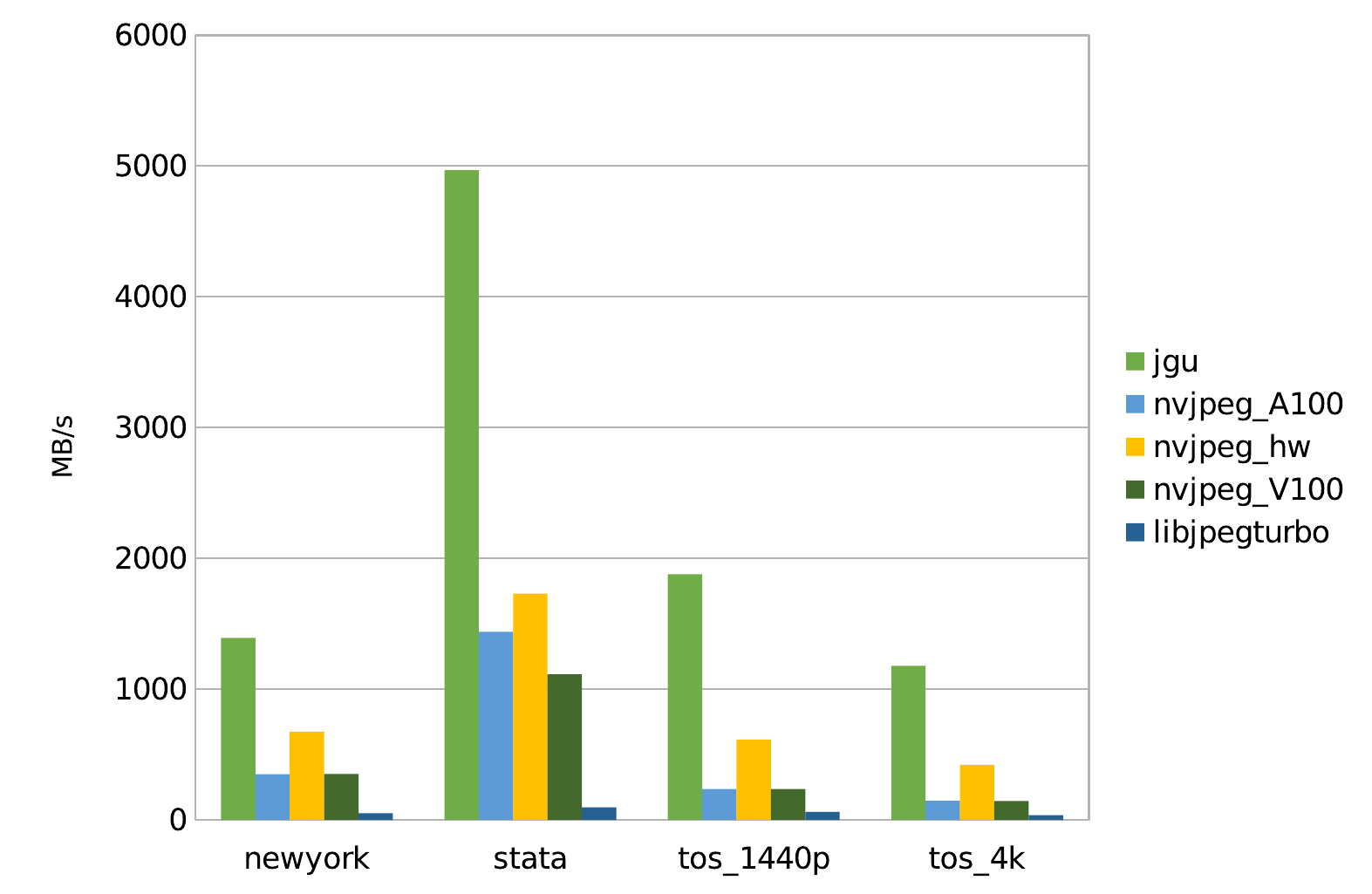}
    \caption{Throughput rates for various datasets}
    \label{fig:throughput_various}
\end{figure}

\begin{figure}[ht]
    \centering
    \includegraphics[scale=0.52, trim= 0 5 0 5, clip=true]{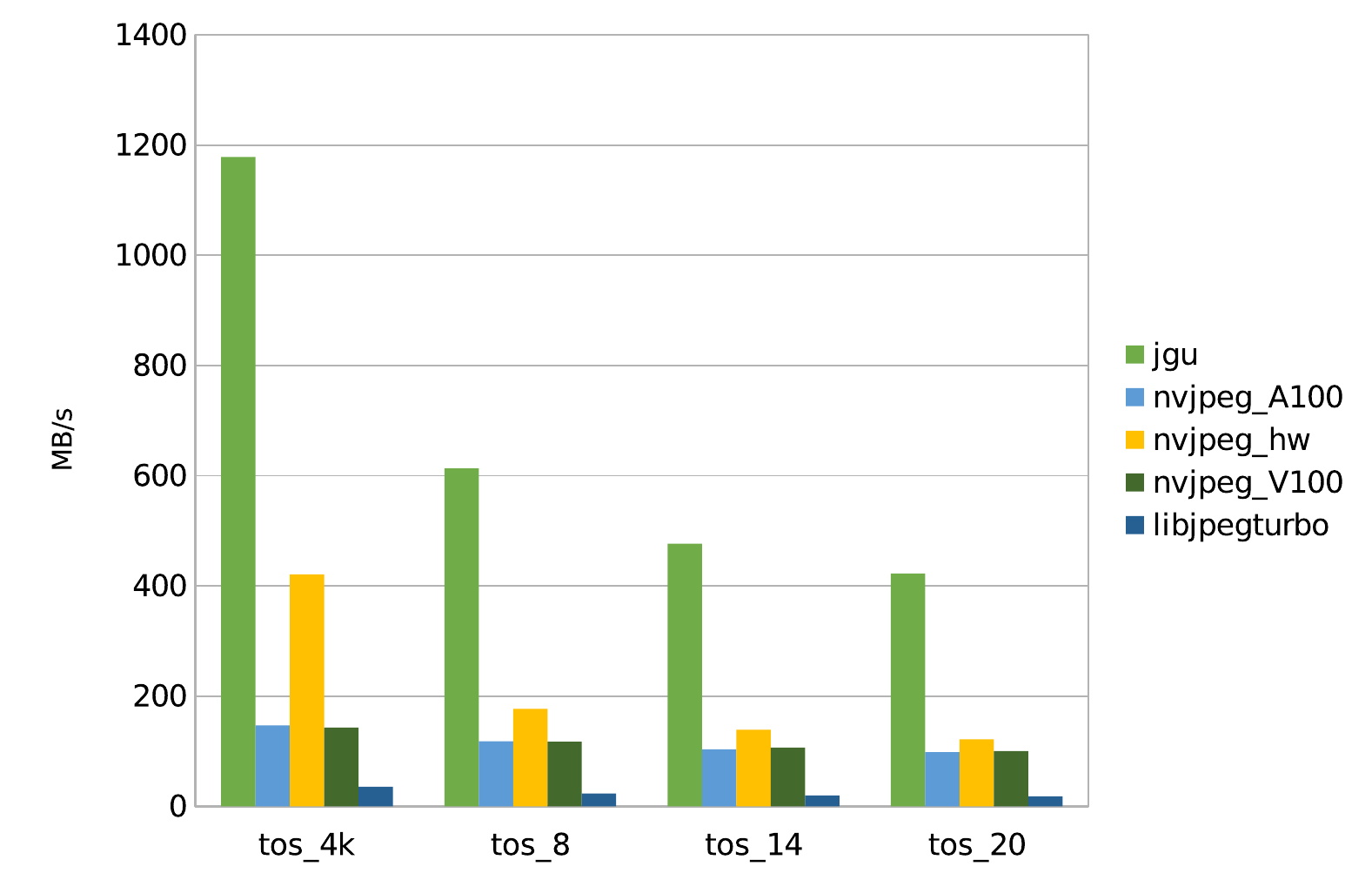}
    \caption{Throughput rates for images of various qualities}
    \label{fig:throughput_quality}
\end{figure}

Throughput rates are presented in Figures \ref{fig:throughput_various} and \ref{fig:throughput_quality}. In these tests, the ‘jgu’ decoder was run exclusively on the A100 GPU. The individual values represent the throughput of \textit{compressed} data per second. ‘jgu’ achieves the highest throughput for each dataset. The highest rate for each tested decoder is achieved for the stata set due to its shorter bitstreams and high image count, allowing for a higher number of images to be processed in parallel by the GPU-based libraries. For stata, the highest throughput of $4967$ MB/s is achieved by ‘jgu’, while ‘nvjpeg\_hw’ is second fastest decoding at a rate of $1729.5$ MB/s. For the newyork dataset, $1390.6$ MB/s and $674.2$ MB/s can be observed, respectively. $1878.2$ MB/s and $1178.2$ MB/s have been achieved by ‘jgu’ for the tos datasets, while ‘nvjpeg\_hw’ has a throughput of $614.8$ and $420.8$, respectively.

Figure \ref{fig:throughput_quality} presents throughput rates for varying image qualities. Considering all libraries, the results reveal deteriorating performance at lower quality. As low image quality implies higher compression, the bitstreams of the individual images are shorter. Thus, the libraries have to spend less time on Huffman decoding, while the time consumed by the rest of the pipeline remains constant. For this reason, more work is performed per bit, reducing the throughput with respect to the compressed data. For 'jgu’, it decreases from the highest quality (tos\_4k) at $1178.2$ MB/s, to $422.3$ MB/s for ‘tos\_20’. For ‘nvjpeg\_hw’, the throughput decreases from $420.9$ MB/s to $121.6$ MB/s. Furthermore, non hardware-accelerated nvJPEG exhibits almost equal performance when run on the A100 and V100 GPUs. libjpegturbo achieves throughputs between $35.4$ and $18$ MB/s.

\section{Conclusion and Future Work}\label{sec:conclusion}

For many GPU-accelerated
computer vision and deep learning tasks,  efficient JPEG decoding is essential due to limitations in memory bandwidth. However, transferring decoded image data to accelerators via slow interconnects such as PCI Express leads to decreased throughput rates which motivates the design of GPU-based JPEG decoders. Previous approaches to this problem have not achieved high efficiency since they often employ hybrid designs where important parts of the decoding pipeline are still executed on the CPU. In this paper, we have presented an approach to parallel JPEG decoding entirely on GPUs based on the self-synchronizing property of Huffman codes. We have implemented our algorithm using CUDA and evaluated its performance on Volta and Ampere-based hardware platforms. 
The results demonstrate that on an A100 (V100) GPU our implementation can outperform the state-of-the-art implementations libjpeg-turbo (CPU) and nvJPEG (GPU) by a factor of up to 51 (34) and 8.0 (5.7). Furthermore, on an A100 it is able to outperform nvJPEG accelerated with dedicated hardware cores by a factor of up to 3.4. Integration of the algorithm in existing deep learning pipelines, e,g. NVIDIA DALI, as well as support for multi-batch processing on multiple GPUs is subject of future work. The IDCT and dequantization stage now represents a bottleneck in the decoding pipeline, thus requiring further research. Also, the results suggest that probabilistic methods based on the overflow pattern could provide fine-grained parallelization of various other tasks in the field of data compression and decompression.

\printbibliography

\end{document}

%% file: alg_decoding.tex
\begin{algorithm}
\SetKwInOut{Input}{Input}
\SetKwInOut{Output}{Output}
\SetKwFunction{Decode}{decode\_subseq}
\SetKwFunction{Sync}{\_\_syncthreads}
\SetKwFunction{syncdecoders}{sync\_decoders}

\SetKwFunction{decsub}{decode\_subsequence}

\SetKwBlock{DoParallel}{do in parallel}{end}

\Input{Buffer \textit{batch} of JPEG bitstreams}
\Output{Buffer \texttt{out} with decompressed bitmap data for all images in \textit{batch}}
\BlankLine
\texttt{out} $\longleftarrow$ allocate output buffer;\\
divide bitstream into equally sized subsequences of $s\cdot 32$ bits; \\
$N \longleftarrow$ number of resulting subsequences; \\
$B \longleftarrow$ $\lceil N / b \rceil$; \\
\texttt{s\_info} $\longleftarrow$ allocate buffer for synchronization;\\
\BlankLine
\syncdecoders{batch}; \\
\BlankLine
\textbf{in parallel} compute exclusive prefix sum on \texttt{s\_info}$.n$;\\
\texttt{s\_info}$[0].n \longleftarrow 0$;\\
\BlankLine
\textbf{for each} subsequence $s_i \in \{0,\dots,b-1\}$ \DoParallel{
    \uIf{$s_i == 0$}{
    \decsub{$s_i$, false, true, \texttt{out}, \texttt{s\_info}};
    }
    \Else{\decsub{$s_i$, true, true, \texttt{out}, \texttt{s\_info}};}
}
\BlankLine
\textbf{for each} \textit{image} in \textit{batch} \DoParallel{
reverse difference coding for all DC coefficients in every component in \textit{image};
}

\textbf{for each} \textit{image} in \textit{batch} \DoParallel{
reverse zig-zag encoding, dequantize and compute the IDCT of all data units in \textit{image};
}

\caption{Decoding of JPEG images (batched)}
\label{alg:write_output}
\end{algorithm}

%% file: alg_single.tex
\begin{algorithm}
\SetKwFunction{Decode}{decode\_subseq}
\SetKwFunction{Sync}{\_\_syncthreads}

\SetKwFunction{decsub}{decode\_subsequence}
\SetKwProg{Fn}{function}{}{}

\SetKwFunction{decsym}{decode\_next\_symbol}
\Fn{\decsub{$i$, \textit{overflow}, \textit{write}, \texttt{out}, \texttt{s\_info}}}{
    $ (p, n, c, z)  \longleftarrow (\textnormal{start of i-th subsequence}, 0, \texttt{Y}, 0)$;\\
    \textit{last\_symbol};\\
    \textit{position\_in\_output} $\longleftarrow 0$;\\
    \BlankLine
    
    \If{\textit{write}}{\
    \textit{position\_in\_output} $\longleftarrow \textnormal{\texttt{s\_info}[i]}.n$;}
    \BlankLine
    
    \If{overflow}{$ (p, n, c, z)  \longleftarrow$ \texttt{s\_info[$i-1$];}}
    \While{not end of subsequence $i$ reached}
    {
        \textit{last\_symbol} $\longleftarrow (p, n, c, z)$; \\
        (\textit{length}, \textit{symbol}, \textit{run\_length}) $\longleftarrow$ \decsym{}; \\
        \If{\textit{write}}{
            \texttt{out}[\textit{position\_in\_output}] = \textit{symbol};
        }
        \textit{position\_in\_output} $\longleftarrow$             \textit{position\_in\_output} + run\_length  + 1;\\
        $p \longleftarrow p + \textnormal{\textit{length}}$;\\
        $n \longleftarrow n + \textnormal{\textit{run\_length}} + 1$; \\
        $z \longleftarrow z + \textnormal{\textit{run\_length}} + 1$; \\
        \If{z $\geq$ 64 or \textit{symbol} == \textbf{EOB}}{$z \longleftarrow 0$;}
        \If{component complete}{$c \longleftarrow$ next component;}
    }
    \textbf{return} \textit{last\_symbol};
    }
\caption{Decoding of a single subsequence in a JPEG scan}
\label{alg:subsequence}
\end{algorithm}

%% file: alg_sync.tex
\begin{algorithm}
\SetKwInOut{Input}{input}
\SetKwInOut{Output}{output}
\SetKwFunction{Decode}{decode\_subseq}
\SetKwFunction{Sync}{\_\_syncthreads}
\SetKwFunction{decsub}{decode\_subsequence}
\SetKwFunction{syncdecoders}{sync\_decoders}
\SetKwProg{Fn}{function}{}{}
\SetKwBlock{DoParallel}{do in parallel}{end}

\Fn{\syncdecoders{batch}}{{
\textbf{for each} \textit{image} in \textit{batch} \DoParallel{
\textit{last } $\longleftarrow$ index of last subsequence in bitstream;\\
\texttt{sequence\_synced} $\longleftarrow$ \texttt{[false$,\dots,$false]};
\textbf{for each} block $b_i \in \{0,\dots,B-1\}$ \DoParallel{
\textbf{for each} subsequence $s_i \in \{0,\dots,b-1\}$ \DoParallel{
\textit{seq\_global} $\longleftarrow$ $b_i \cdot \,b$; \\
\textit{subseq\_global} $\longleftarrow$ $b_i \cdot \,b + s_i$; \\
\textit{synchronized} $\longleftarrow$ \textit{false}; \\
\textit{end} $\longleftarrow \min$(\textit{seq\_global} $+\,b$, \textit{last});\\

\texttt{s\_info}[\textit{subseq\_global}] $\longleftarrow$ \decsub{subseq\_global, false, false, \texttt{out}, \texttt{s\_info}};\\
\Sync{};\\
++\textit{subseq\_global};\\
    \While{not synchronized and subseq\_global $\leq$ end}{
        $(p, n, c, z) \longleftarrow$ \decsub{\\subseq\_global, true, false, \texttt{out}, \texttt{s\_info}};\\
        \If{$(p, c, z) ==$ \texttt{s\_info}[\textit{subseq\_global}]}{
            \textit{synchronized} $\longleftarrow$ \textit{true};\\
        }
        \texttt{s\_info} $\longleftarrow (p, n, c, z)$;\\
        ++\textit{subseq\_global};\\
        \Sync{}\;
    }
}
}
\While{$\exists$ unset flag in \texttt{sequence\_synced}}{
\textbf{for each} block $b_i \in \{0,\dots,B-2\}$ \DoParallel{
\textit{subseq\_global} $\longleftarrow$ $(b_i + 1) \cdot \,b$; \\
\textit{synchronized} $\longleftarrow$ \textit{false}; \\
\textit{end} $\longleftarrow \min$(\textit{subseq\_global} $+\,b$, \textit{last});\\
    \While{not synchronized and subseq\_global $\leq$ end}{
        $(p, n, c, z) \longleftarrow$ \decsub{\\subseq\_global, true, false, \texttt{out}, \texttt{s\_info}};\\
        \If{$(p, c, z) ==$ \texttt{s\_info}[\textit{subseq\_global}]}{
            \textit{synchronized} $\longleftarrow$
            \textit{true};\\
            \texttt{sequence\_synced}[$b_i - 1$] $\longleftarrow$ \textit{true};\\
        }
        \texttt{s\_info} $\longleftarrow (p, n, c, z)$;\\
        ++\textit{subseq\_global};\\
        \Sync{}\;
    }
}
}
}
}
}
\caption{Synchronizing decoders (JPEG, batched)}
\label{alg:decoder}
\end{algorithm}